\documentclass[aps,prl,reprint,groupedaddress]{revtex4-1}

\usepackage{float}
\usepackage{hyperref}
\usepackage{epsfig}
\usepackage{graphicx}
\usepackage{subfigure}
\usepackage{latexsym}
\usepackage{color}
\usepackage{fullpage}
\usepackage{epstopdf}
\usepackage{dcolumn}
\usepackage{bm}
\usepackage{ulem}
\usepackage{units}
\usepackage{amsmath}
\usepackage{lineno}
\usepackage{verbatim}

\newcommand{\nico}[1]{\textcolor{black}{#1}}

\begin{document}

\title{The square ice Coulomb phase as a percolated vertex lattice}

\author{J. Coraux}
\email{johann.coraux@neel.cnrs.fr}
\author{N. Rouger}
\author{B. Canals}
\author{N. Rougemaille}
\email{nicolas.rougemaille@neel.cnrs.fr}
\affiliation{Université Grenoble Alpes, CNRS, Grenoble INP, Institut NEEL, 38000 Grenoble, France}

\begin{abstract}
The square ice is a canonical example of a Coulomb phase in two dimensions: its ground state is extensively degenerate and satisfies a local constraint on the spin arrangement (the so-called ice rule).
In this work, we use a loop flip algorithm to explore the properties of this ground state that we analyze not in terms of a spin texture, but rather in terms of a spatial distribution of ice-rule satisfying vertices.
More specifically, we determine for various lattice sizes the average vertex populations characterizing the ice manifold, the pairwise vertex correlations and the size distribution of vertex clusters.
Comparing these results to those obtained from random, constraint-free vertex tilings, the square ice manifold is found to resemble an almost ideal vertex gas, and the cluster size distribution of ice-rule satisfying vertices is well approximated by percolation theory.
Remarkably, this description remains reasonably accurate when monopoles are present in dilute amount, allowing direct comparison with experiments. 
Revising former experimental results on two artificial square ice systems, we illustrate the interest of our approach to spot the presence of a Coulomb phase from a vertex analysis.
\end{abstract}

\maketitle

\section{I. Introduction}

\begin{figure}
\begin{center}
\includegraphics[width=58.5664 mm]{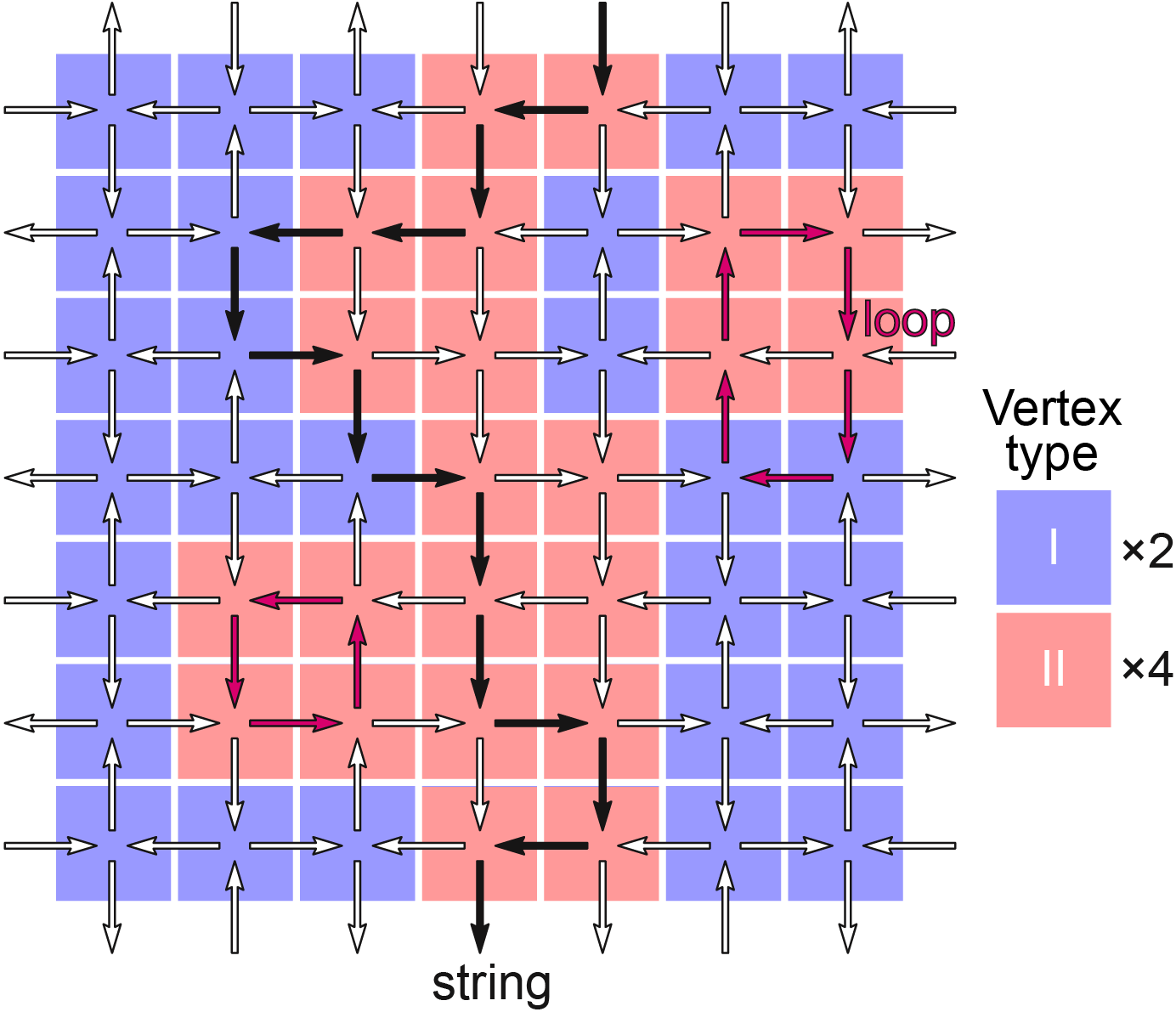}
\caption{\label{fig1} Ising spins (arrows) arranged and oriented along the bonds of a square lattice with open boundary conditions. Examples of two loops and a string spanning across the lattice are highlighted in pink and black. Flipping the spin directions within such loops or strings allows the system to fluctuate without breaking the two in / two out local constraint. The vertex map, associated with the spin configuration, is shown as a colored background, with type I and type II vertices appearing in blue and red, respectively. Numbers indicate the vertex degeneracy.}
\end{center}
\end{figure}

The square ice \cite{Lieb1967a, Lieb1967b} is a frustrated magnet.
It is defined by a set of Ising variables, placed and oriented along the bonds of a two-dimensional square lattice in such a way that each vertex is made of two spins pointing inward and two spins pointing outward (Fig.~\ref{fig1}). 
The square ice manifold is extensively degenerate and thus characterized by a residual entropy per site \cite{Lieb1967b}. 
This manifold consists of all tilings based on the six possible vertex arrangements satisfying the two in / two out constraint (see the type I and type II vertices shown in Fig.~\ref{fig1} with their associated degeneracy).
This contraint is the ice rule introduced by Bernal and Fowler \cite{Bernal1933} in water ice \cite{Pauling1935}, which translates into a local divergence-free condition of the magnetization field vector.
Fluctuating from one ice micro-state to another one cannot be achieved by flipping a single spin, which necessarily violates the ice rule constraint, generating a pair of magnetically charged defects, the so-called magnetic monopoles \cite{Ryzhkin2005, Castelnovo2008}. 
Fluctuations within the ice manifold then require other spin flip processes preserving the ice rule constraint.

The square ice provides one exemple, among others in the larger family of spin ice systems \cite{Harris1997, Gingras, Bramwell2013, Bramwell2020}, and in highly frustrated magnets on a broader perspective \cite{bookLacroix}, of a freezing of the single spin flip dynamics before the ground state can be reached.
Its low-energy properties can be probed using collective spin updates, for example along either closed loops, fully comprised within the lattice or wrapping the torus under periodic boundary conditions (PBC), or strings, i.e., portions of loops intersected by the lattice edges under open boundary conditions (OBC) (see Fig.~\ref{fig1}).
In fact, loop updates play a key role in spin ice physics, and they are intimately related to the concept of Coulomb phase \cite{Henley2010, Moessner2016}.
Although the concept is quite general and applies to many types of lattice models, whether they are magnetic or not, frustrated magnets are probably the most known systems in which the Coulomb phase physics is actively studied \cite{Brooks2014, Elsa2020, Benton2021, Pujol2023, Benton2023}. 
Coulomb phases have been also investigated in square \cite{Perrin2016, Ostman2018, Farhan2019, Brunn2021, Goryca2021, Rougemaille2021, Schanilec2022} and kagomé \cite{Canals2016, Sendetskyi2016, Schanilec2020, Yue2022, Hofhuis2022} artificial spin ices, in which their properties can be visualized in real space and time, at the scale of the individual spin degree of freedom \cite{Nisoli2013, Rougemaille2019, Skjaervo2020}.

\section{II. Motivation}

The question we address in this work is whether the Coulomb phase that characterizes the square ice presents peculiar signatures when analyzed not in terms of a magnetization texture but rather in terms of a vertex distribution. 
In other words, we wonder whether the square ice Coulomb phase has signatures other than peculiar magnetic correlations.
As we will see, the answer to this question is yes, and we report below three main results.
1) The square ice manifold is well approximated by an unconstrained random distribution of type I and type II vertices when their fractions are set to 37$\%$ and 63$\%$, respectively, for large lattices. 
Because of the large fraction of type II vertices, the vertex distributions within the square ice are well described by those of a percolated square lattice, whose properties are known in two dimensions. 
2) This description remains reasonable as long as the vertex populations are close to these fractions with a several $\%$ tolerance. 
Interestingly, it also remains relatively accurate when monopoles are present in dilute amount (up to several $\%$, typically), enabling direct comparison with experiments.
3) We confront two experimental case studies from previous literature to our numerical results, and analyse to what extent the picture of an almost ideal gas of ice-rule satisfying vertices accounts for the experiments.

\begin{figure}
\begin{center}
\includegraphics[width=7.8 cm]{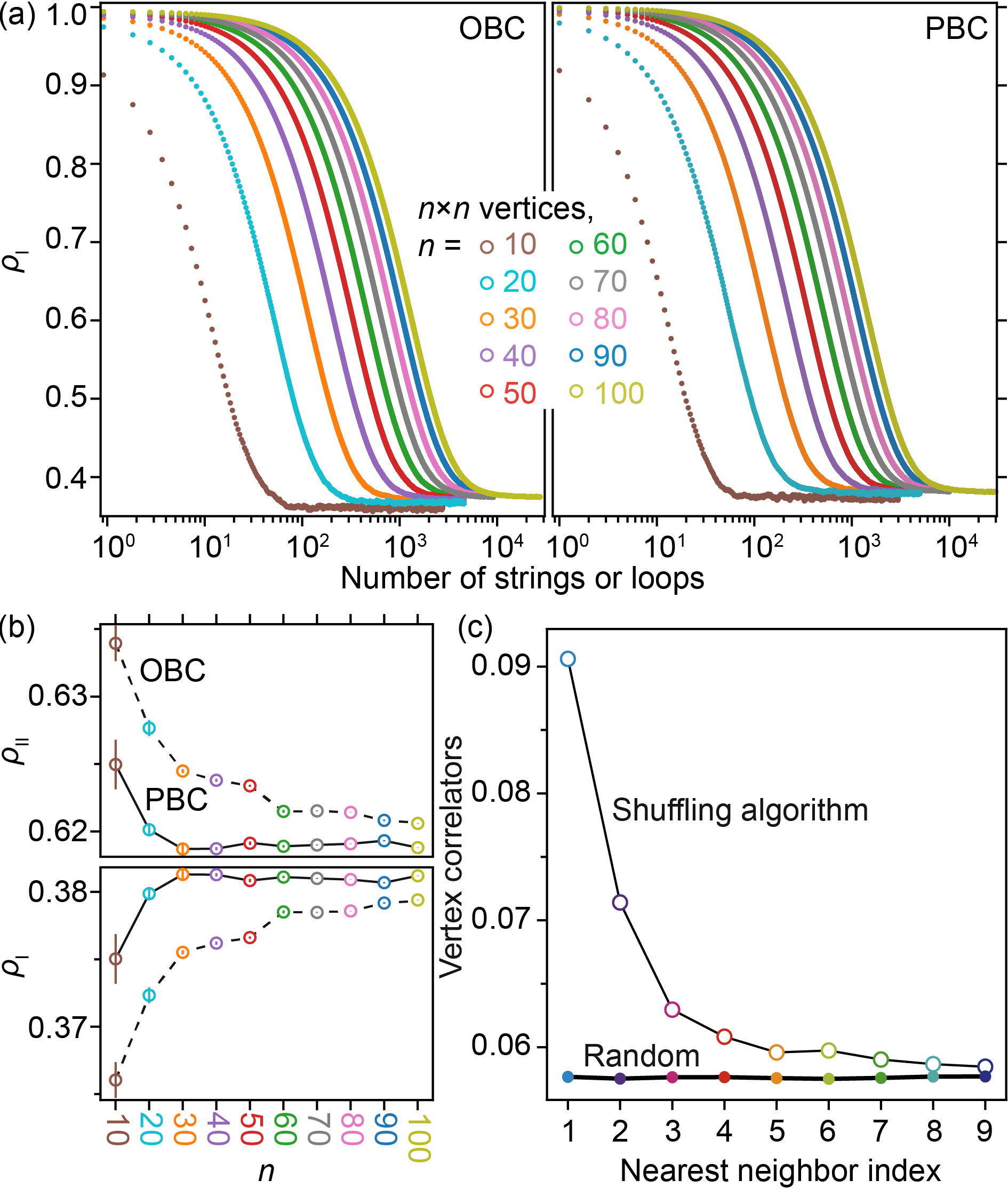}
\caption{\label{fig2} (a) Fraction $\rho_I$ of type I vertices for different lattice sizes ($n\times n$ vertices) against the number of loops and strings that were reversed in the shuffling algorithm. Similar results are obtained for open (OBC) and periodic (PBC) boundary conditions. (b) Limit values of $\rho_I$ and $\rho_{II}$ (fractions of type I and type II vertices, respectively) averaged over $N=10^3$ configurations for various lattice sizes. The same color code is used on both figures. The standard deviation is represented by the vertical colored lines (very small for $n\geq30$). (c) Pairwise vertex correlations as a function of the separation distance computed from the shuffling algorithm and from the random tiling approach ($n=100$, OBC). Standard deviation is $\sim10^{-5}$.}
\end{center}
\end{figure}

\section{III. Vertex populations, vertex correlations and cluster size distribution in the square ice}

The numerical approach we followed consists in generating a set of ground state configurations using an algorithm shuffling spins along oriented loops.
We proceeded as follow: 
A starting configuration satisfying the ice rule constraint (in practice the ordered antiferromagnetic state made of a perfect tiling of type I vertices, \nico{but we checked that the initial condition does not modify the results}) is shuffled by flipping a sufficiently large number of randomly chosen (closed) loops and strings (touching two of the system boundaries, see Fig.\ref{fig1}). 
As shown in Fig.~\ref{fig2}(a), the fraction $\rho_I$ of type I vertices continuously decreases from 1 to a (size dependent) limit value after $n^2$ loops and strings have been flipped, $n^2$ being the number of vertices in the lattice. 
We note that the efficiency of the shuffling process does not depend on the boundary conditions, and similar results are found whether the system has open or periodic boundary conditions [Fig.~\ref{fig2}(a)].
The process is repeated $N$ times to obtain a set of configurations belonging to the ground state manifold. 
For each of the $N$ configurations, a vertex map is generated. 
These $N$ maps are then used to compute the vertex populations, the vertex-vertex correlations, and the size distribution of the clusters formed by type I and type II vertices. 
These quantities are also computed for vertex lattices generated in a different way by choosing each vertex randomly (i.e., disregarding the ice rule, imposing no constraint between neighboring vertices), thus generating a $n\times n$ random tiling of the two vertex types.
The results of the two methods are then compared. 

The vertex populations obtained with the shuffling algorithm are displayed in Fig.~\ref{fig2}(b) for different lattice sizes. 
These populations marginally depend on the lattice size, and tend to constant values for the largest sampled arrays. 
Note that the limit values for these two populations are 38\% and 62\% for type I and type II vertices, respectively, and not 1/3 and 2/3 as one might guess from the vertex degeneracy.
In fact, these 1/3 and 2/3 values correspond to the values expected if the vertices were independent, which is not the case here because of the constraint that links two neighboring vertices by a common spin. 
We also note that the values found for large lattices do not depend substantially on the boundary conditions (open or periodic).

\begin{figure}[!hbt]
\begin{center}
\includegraphics[width=7.8 cm]{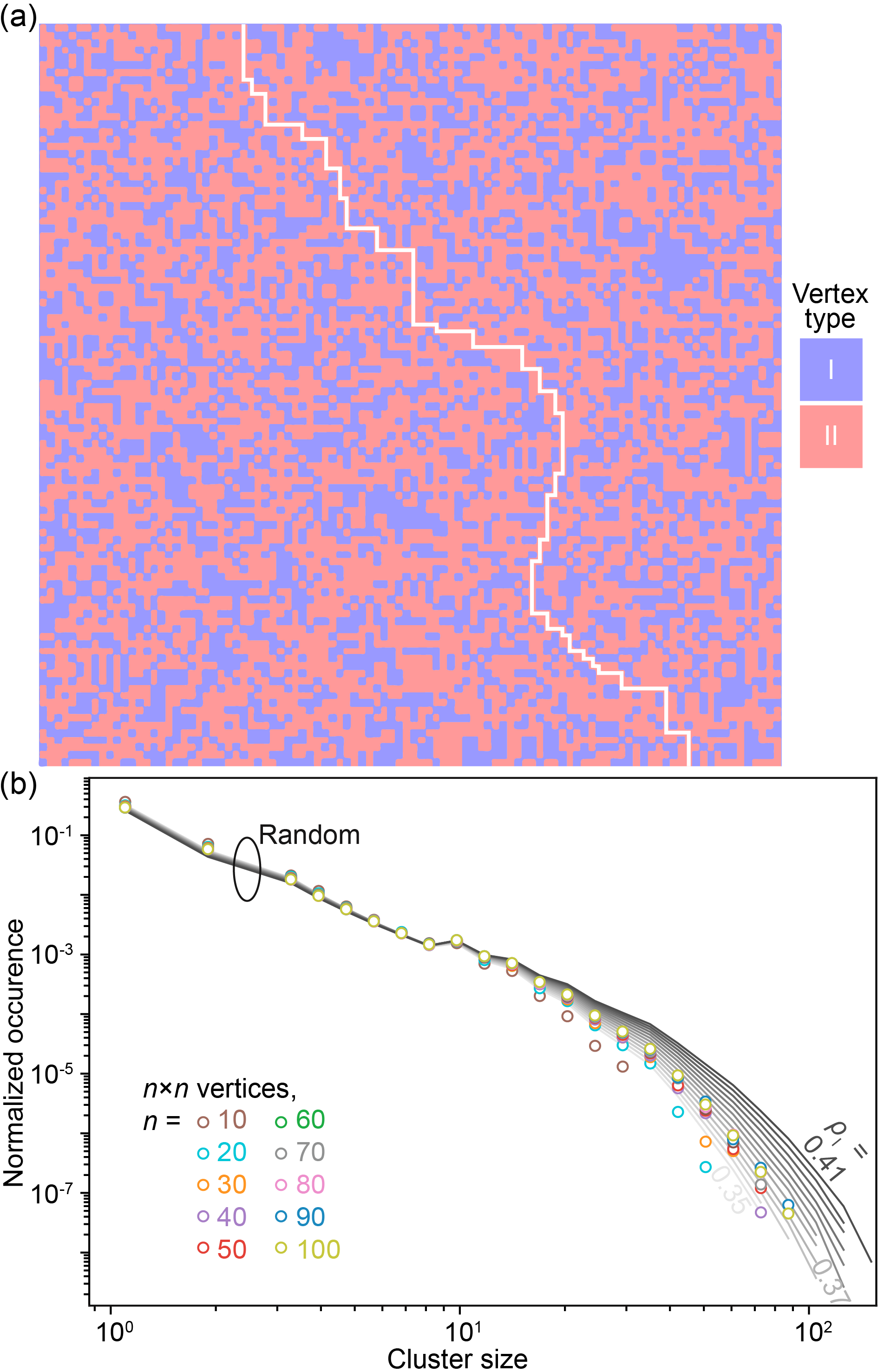}
\caption{\label{fig3}(a) Typical vertex map obtained from the shuffling algorithm. Type I and type II vertices appear in blue and red, respectively. Due to their large fraction (62$\%$), type II vertices percolate. A percolated path is highlighted by a white line. (b) Statistics of the type I cluster size distribution for several lattice sizes ($n\times n$ vertices) under open boundary conditions. Data is averaged over 10$^3$ micro-states, except for the 100$\times$100 lattice for which 4$\times$10$^4$ micro-states have been used. The case of a 100$\times$100 random lattice of type I/II vertices with varying population of type I vertices is also plotted for comparison (for better readability solid lines are used, yet the data were obtained for the same domain size binning as for the shuffling algorithm; the $\rho_\mathrm{I}$ value, between 0.35 and 0.41, is coded by the gray shade, with $\Delta\rho_\mathrm{I}=0.05$ increments).}
\end{center}
\end{figure}

The vertex maps generated with the shuffling algorithm allow us to determine the vertex-vertex correlators in the Coulomb phase manifold. 
Since the vertices are either of type I or type II in the ground state, they can be assigned an Ising variable, and the correlation between neighboring vertices is simply expressed by the quantity $\left\langle \sigma_i \sigma_j \right\rangle$ averaged over all $(i,j)$ vertex pairs, with $\sigma_i=\pm1$ depending on the vertex type.
We can then compute the value of these correlators as a function of the vertex separation distance and compare the results from the shuffling algorithm and the random tiling approach (for which we naturally choose 38$\%$ / 62$\%$ fractions for the two vertex types). 
As anticipated, both methods return small values of a few $\%$ only [Fig.~\ref{fig2}(c)].
At relatively long distances, all correlator values tend to a constant of about 6$\times 10^{-2}$, which corresponds to the value obtained from the random tiling approach: The vertex correlations in the square ice are then very similar to those characterizing an ideal vertex gas.
We emphasize that, although the correlators differ in the two approaches at short distances, the values are fairly small, making them also potentially difficult to distinguish if the statistics is low, like it is often the case experimentally (see below).
Here as well, the results do not depend much on the boundary conditions.

Another important information that can be extracted from the vertex maps obtained with the two methods is the size distribution of clusters made of edge-connected vertices of a given type (I or II). 
These clusters have size distributions (expressed in number of vertices per cluster) that depend on the fraction $\rho$ of the considered vertex type.
Given the percolation threshold $\rho_\mathrm{C}\simeq0.593$ for a square lattice (randomly filled with two different types of edge-touching square elements), the distributions of type I and type II vertices are expected to distribute rather differently since $\rho_\mathrm{I}<\rho_\mathrm{C}$ and $\rho_\mathrm{II}>\rho_\mathrm{C}$.
This can be seen in Fig.~\ref{fig3}(a) where a typical vertex map is reported, showing a large type II cluster that extends throughout the lattice and bridges facing edges [see white path in the figure].
In the following, we will only consider the cluster distribution of type I vertices to characterize the ground state manifold of the square ice.

First, we find a striking similitude of the cluster size distributions obtained with the two algorithms [compare the colored dots and the gray curves in Fig.~\ref{fig3}(b)].
This demonstrates numerically a simple result: a percolation problem on a square lattice accounts well for the vertex distribution in the square ice manifold. 
While this result may seem intuitive, one should keep in mind that a randomly tiled vertex lattice is a priori different from a Coulomb phase.
Indeed, certain configurations present in an unconstrained random arrangement of vertices do not exist in the square ice due to the spin constraint linking neighboring vertices.
For example, a type II vertex surrounded by eight type I vertices cannot be found in a Coulomb phase. 
Despite this fundamental difference, our result shows that the square ice Coulomb phase is an almost ideal gas of ice-rule satisfying vertices.

In fact, differences between the two distributions become discernable for large cluster sizes: For example, as show in Fig.~\ref{fig3}(b), the distribution of type I vertices in a 100$\times$100 lattice slightly deviates from the 38$\%$/62$\%$ random distribution, and replacing the initial 38$\%$/62$\%$ fractions by 37$\%$/63$\%$ gives a much better agreement. 
In other words, the local constraint present in the spin model translates into a change of about 1$\%$ in the vertex fractions used in the random tiling approach.
This is also observed when the calculations are run in periodic boundary conditions (data not shown). 
In all cases and whatever the lattice size, $\rho_\mathrm{I}$ fractions smaller than 38\% in the random tiling approach better account for the results found with the shuffling algorithm.

\begin{figure}
\begin{center}
\includegraphics[width=7.8 cm]{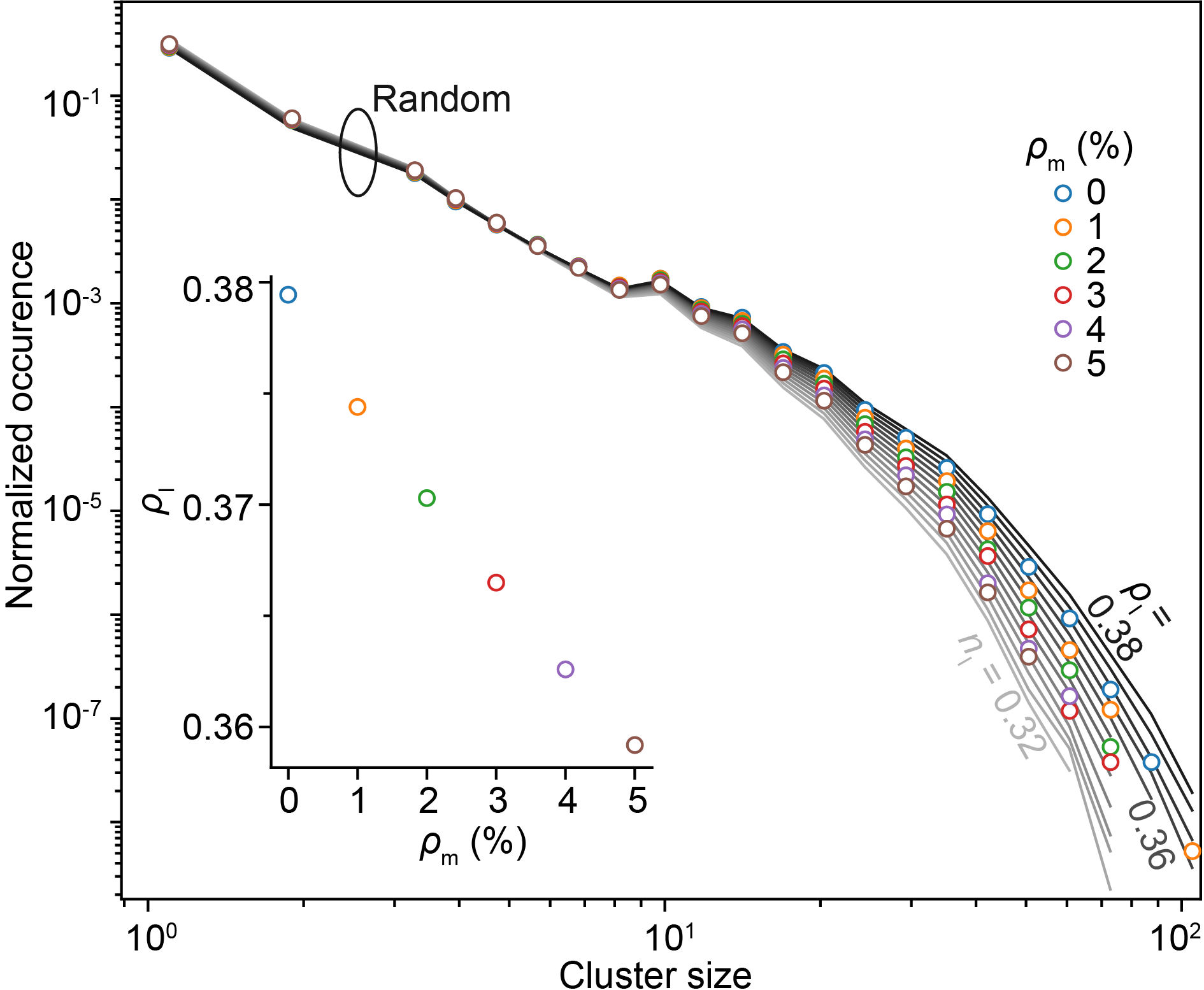}
\caption{\label{fig4} Statistics of the type I cluster size distributions for the 100$\times$100 lattice under open boundary conditions and for several monopole densities $\rho_\mathrm{m}$. Data is averaged over 2$\times 10^3$ micro-states. The case of a random tiling of type I/II vertices (no monopole are present) with varying populations of type I vertices is plotted for comparison (for better readability solid lines are used, yet the data were obtained for the same domain size binning as for the shuffling algorithm; the $\rho_\mathrm{I}$ value, between 0.32 and 0.38, is coded by the gray shade, with $\Delta\rho_\mathrm{I}=0.05$ increments). Inset shows how the fraction of type I vertices varies in the shuffling algorithm as monopoles are injected in the lattice.}
\end{center}
\end{figure}

For relatively small cluster size $s$, the $\rho_\mathrm{I}$ distribution from the shuffling algorithm is well described by a power law \cite{Stauffer1992, Christensen2005}.
For large $s$ values, $\rho_\mathrm{I}$ decreases faster, which may altogether be accounted for by multiplying the at-percolation power-law with an ad hoc function of $s$, for instance an exponential \cite{Ding2014}.

Summarizing our results, one can consider that a spin configuration presenting a vertex fraction of the order of 38$\%$ / 62$\%$, a first-neighbor vertex correlator of the order of 0.06, and a size distribution of type I clusters described by the above-mentioned power law to be a Coulomb phase micro-state. 
We stress that the magnetic correlations of the underlying spin texture are not considered here, and only the vertex statistics matters. 

In practice, particularly in artificial spin ice systems where vertex maps can be measured, the ground state is not reached. 
For example, artificial arrays of interacting nanomagnets generally present a nonzero density of monopoles, which we have not considered so far. 
The question is therefore to what extent the description of artificial realizations of the square ice using percolation theory remains a good approximation.
To address this issue, we proceeded the same way as before and compared the outcome of the shuffling algorithm, applied to starting configurations where monopoles have been injected with a constant density $\rho_m$ of 1, 2, 3, 4 and 5$\%$, to the purely random distributions.
We note that the fraction of type I/II vertices reduces linearly as the monopole density increases (see inset of Fig.~\ref{fig4}).
This is expected and a monopole should replace a type I or type II vertex with a 38$\%$ or 62$\%$ probability, in good agreement with what is found.
As monopoles tend to break large clusters into smaller ones, the cluster size distributions feature less large clusters as $\rho_m$ increases.
Remarkably, good agreement is again found with a fully random vertex tiling, provided that the $\rho_\mathrm{I}/\rho_\mathrm{II}$ ratio is decreased accordingly to the density of injected monopoles (by few to several percents in the $\rho_m$ range studied here, see Fig.~\ref{fig4}).
The comparison between the two approaches is sufficiently robust to envisage a practical application to artificial spin ices.

\section{IV. Application to artificial square ice systems}

\begin{table}
\caption{\label{table1}
Vertex populations for type I and type II vertices ($\rho_\mathrm{I}$ and  $\rho_\mathrm{II}$, respectively) as well as monopoles ($\rho_m$), and first neighbor vertex correlator $C_1$ in different artificial spin ice (ASI) structures. The lattice size is $n \times n$ vertices. Populations expected in a random vertex tiling containing a monopole density $\rho_m$ are indicated as $\rho_\mathrm{I}^{gas}$ and $\rho_\mathrm{II}^{gas}$ for type I and type II, respectively \cite{note_table}.
 }
\begin{tabular}{|p{1cm}|p{0.5 cm}|p{0.85 cm}|p{0.85 cm}|p{0.85 cm}|p{0.85 cm}|p{0.85 cm}|p{0.85 cm}|}
\hline
\hline
 $lattice$ & $n$ & $\rho_\mathrm{I}$ & $\rho_\mathrm{II}$ & $\rho_m$ & $C_1$ & $\rho_\mathrm{I}^{gas}$ & $\rho_\mathrm{II}^{gas}$ \\
\hline
1 \cite{Perrin2016} & 20 & 0.285 & 0.590 & 0.125 & 0.09 & 0.322 & 0.553 \\
\hline
2 \cite{Perrin2016} & 20 & 0.320 & 0.598 & 0.083 & 0.14 & 0.339 & 0.578 \\
\hline
3 \cite{Perrin2016} & 20 & 0.280 & 0.615 & 0.105 & 0.14 & 0.330 & 0.565 \\
\hline
4 \cite{Perrin2016} & 20 & 0.283 & 0.598 & 0.120 & 0.08 & 0.324 & 0.556 \\
\hline
5 \cite{Perrin2016} & 20 & 0.320 & 0.600 & 0.080 & 0.12 & 0.340 & 0.580 \\
\hline
6 \cite{Perrin2016} & 20 & 0.310 & 0.618 & 0.073 & 0.10 & 0.343 & 0.584 \\
\hline
7 \cite{Perrin2016} & 20 & 0.298 & 0.625 & 0.078 & 0.19 & 0.341 & 0.581 \\
\hline
8 \cite{Perrin2016} &20 & 0.233 & 0.678 & 0.090 & 0.21 & 0.336 & 0.574 \\
\hline
9 \cite{Schanilec2022} & 30 & 0.333 & 0.663 & 0.003 & 0.27 & 0.374 & 0.623 \\
\hline
10 \cite{Schanilec2022} & 30 & 0.327 & 0.672 & 0.001 & 0.21 & 0.375 & 0.624 \\
\hline
\hline
11 \cite{Schanilec2022} & 30 & 0.460 & 0.533 & 0.007 & 0.21 & 0.372 & 0.621 \\
\hline
12 \cite{Schanilec2022} & 30 & 0.477 & 0.522 & 0.001 & 0.17 & 0.375 & 0.624 \\
\hline
\hline
\end{tabular}
\end{table}

\begin{figure}
\begin{center}
\includegraphics[width=7 cm]{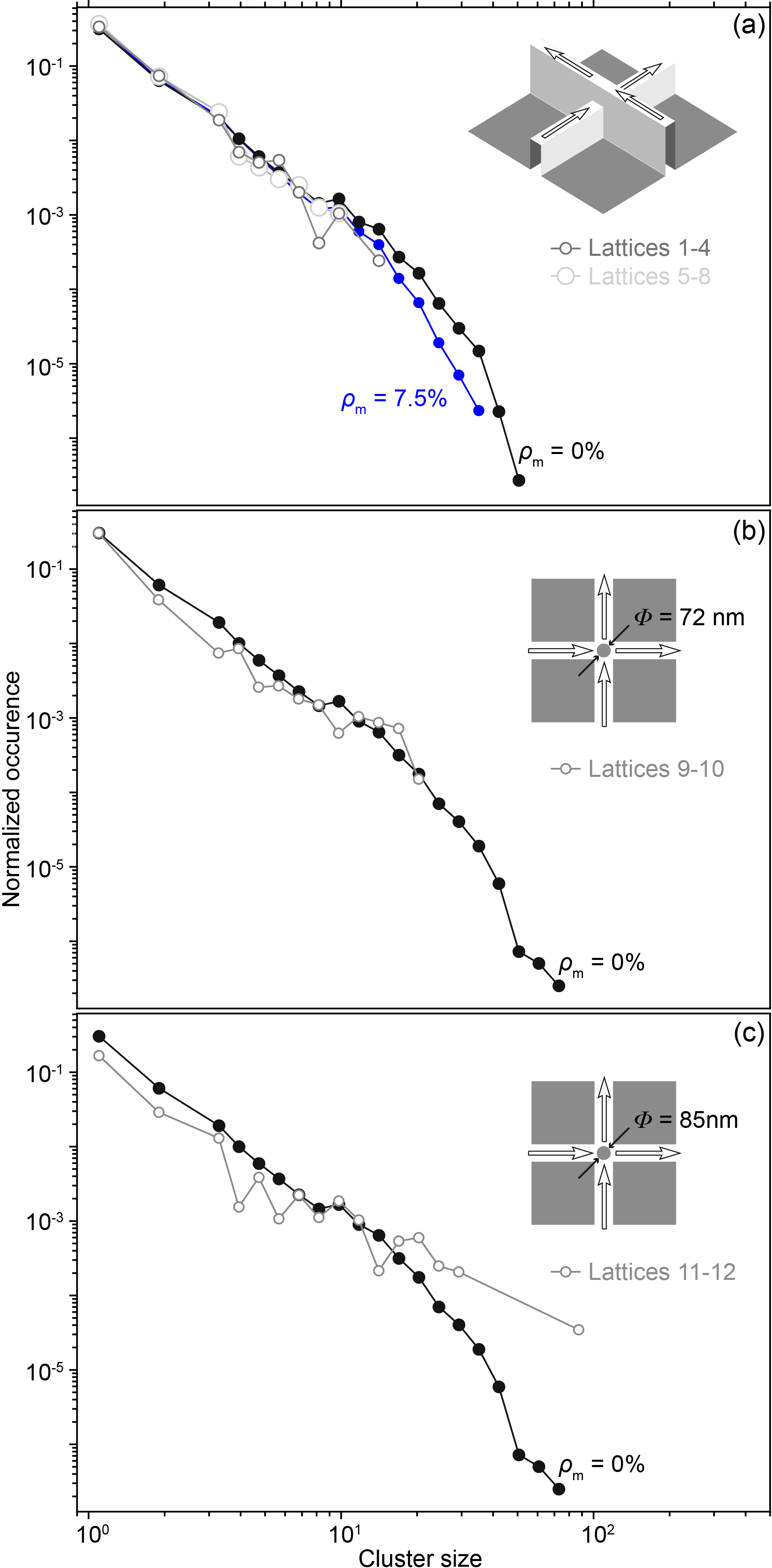}
\caption{\label{fig5} Statistics of the type I cluster size distributions for different experimental situations. (a) Two lattices consisting of 20$\times$20 vertices in which one sublattice is shifted vertically to restore the ice condition between all nanomagnets \cite{Perrin2016}. (b) Two lattices consisting of 30$\times$30 vertices that physically connect the neighboring nanomagnets, but leaving an empty hole at the vertex site to retrieve the type I / type II energy degeneracy \cite{Schanilec2022}. (c) Two lattices consisting of 30$\times$30 vertices like in (b), but in which the ice condition is detuned \cite{Schanilec2022}. In all three cases, the experimental measurements are compared to predictions from the shuffling algorithm. Although good agreement is found for the first two case scenarios (a,b), the model fails to reproduce the results in the third case (c), especially for large cluster sizes. In the insets of (b,c), $\phi$ indicates the diameter of the hole left at the vertex center.}
\end{center}
\end{figure}

We finally consider two experimental situations known to realize of a Coulomb phase physics \cite{Perrin2016, Schanilec2022}. 
We re-examine these previously published results through the analysis of the vertex statistics, following the methodology presented above.
The two cases are interesting as they represent two distinct strategies to realize the square ice experimentally.
The first data set has been obtained by modifying the two-dimensional square ice and exploiting the third direction of space, perpendicular to the lattice plane, to render the first and second neighbor coupling strengths equal \cite{Perrin2016, Brunn2021} [see inset of Fig.~\ref{fig5}(a)]. 
The second data set has been obtained for a truly planar square lattice in which the nanomagnets are physically connected, but with an empty hole at the vertex center to restore the energy degeneracy of type I and type II vertices \cite{Schanilec2022} [see inset of Fig.~\ref{fig5}(b)].
The vertex populations $\rho_\mathrm{I}$ and $\rho_\mathrm{II}$, the monopole density $\rho_m$ and the first neighbor vertex correlator $C_1$ are reported in Table~\ref{table1}. 
We stress that, while the two strategies led to the observation of a Coulomb phase, the table shows that the vertex statistics differ substantially in the two cases. 
In particular, the monopole density varies by almost an order of magnitude between the two realizations \cite{note}. 

The experimentally measured cluster size distributions of type I vertices are compared with those deduced from the shuffling algorithm. 
The results of this comparison are reported in Figs.~\ref{fig5}(a) and ~\ref{fig5}(b) for the two case scenarios. 
Remarkably, good agreement is found, demonstrating that the experimentally measured Coulomb phase micro-states are well described by an almost ideal vertex gas, despite the presence of magnetic monopoles, as well as vertex populations and correlation differing from the expected ones.
In particular, it is interesting to note that the $\rho_\mathrm{I}$ / $\rho_\mathrm{II}$ vertex populations measured in ten artificial square ice realizations are \textit{systematically} lower / larger than the ones expected in a purely random vertex tiling containing a monopole density $\rho_m$ (see Table~\ref{table1}).
In other words, despite the fact that these ten lattices are characterized by magnetic correlations resembling those of the square ice \cite{Perrin2016, Schanilec2022}, our vertex analysis reveals that experimental conditions can be further optimized to truly match the Coulomb phase properties.
For example, the analysis performed on the first eight lattices indicates that the height offset (100 nm, see Ref. \onlinecite{Perrin2016}) used to restore the ice degeneracy in field demagnetized systems might be slightly too large.
Reducing this height offset should provide even better agreement with the square ice physics, consistent with the coupling strength analysis reported in Ref. \onlinecite{Brunn2021}.
The same interpretation could be also made for lattices 9 and 10, for which the hole diameter inserted at the lattice vertices might be slightly too small.
Increasing $\phi$ might bring the system closer to the square ice Coulomb phase, although the fairly large value of the vertex correlator $C_1$ suggests that the vertex distribution could be biased (see Table~\ref{table1}).
Indeed, it is likely that the field demagnetization protocol employed in this work favors the formation of large type II clusters, kinetically stabilized by magnetic domain walls that can propagate throughout the lattice, contrary to what happens in systems in which the nanomagnets are physically disconnected \cite{Schanilec2022}.
  
Finally, we consider other experimental cases known to deviate slightly from the properties of a Coulomb phase \cite{Schanilec2022}.
These systems were obtained from the second strategy by increasing the hole diameter left at the vertex center so that the ice rule condition is detuned.
In the corresponding micro-states, antiferromagnetic correlations are more pronounced than what they should be in the Coulomb phase, and the vertex populations deviate from the populations expected in the ice regime (lattices 11 and 12 in Table~\ref{table1}). 
Comparison with the shuffling algorithm reveals a clear discrepancy, notably for large cluster sizes [see Fig.~\ref{fig5}(c)].
There, the experimental data cannot be faithfully reproduced, and the magnetic configurations cannot be considered as an almost ideal vertex gas.

These different comparisons illustrate the central result of this study: the Coulomb phase of the square ice is well approximated by an almost ideal vertex gas, even when monopoles are present in the system.
Our results also suggest that a phase in which the vertex cluster distribution does not follow the analytical form derived from percolation theory is likely not a Coulomb phase.
The size distribution of the largest clusters in particular can be used as a sensitive 'criterion'.
In the context of artificial spin ice physics, the type I cluster size distribution thus appears as a relevant quantity to spot the presence of a Coulomb phase, complementary to what can be done by computing the magnetic correlations or the magnetic structure factor.
\nico{To conclude, we stress that we have studied experimentally the case of field demagnetized artificial square ice systems.
It would be interesting to proceed the same way with thermally active lattices, such as those reported in Refs. \onlinecite{Ostman2018, Farhan2019}.
However, the monopole fractions is large (about 37\% and 23\%, respectively) in those systems in which the interactions between neighboring nanomagnets is weak due to their small volume.
As a consequence, the fraction of type II vertices is below (or barely reaches) the percolation threshold.
Reducing the density of magnetic monopoles is thus a prerequisite to extend our analysis to thermally active systems.}

\section{acknowledgements}
This work was supported by the Agence Nationale de la Recherche through project no. ANR-22-CE30-0041-01 'ArtMat'.


\begin{thebibliography}{2}
\bibitem{note_table}
The two vertex populations expected in a random vertex tiling containing a monopole density $\rho_m$ are calculated such that $\rho_\mathrm{I}^{gas} = \rho_\mathrm{I}^{limit} - 0.4 \rho_m$ and $\rho_\mathrm{II}^{gas} = \rho_\mathrm{II}^{limit} - 0.6 \rho_m$, where $\rho^{limit}$ is the limit value of the associated vertex fraction for a $20 \times 20$ or $30 \times 30$ lattice size. The 0.4 and 0.6 prefactors are deduced from the linear dependence of $\rho_\mathrm{I}$ against $\rho_m$ (see inset of Fig.~\ref{fig4}).
\bibitem{note}
In the image analysis of the two experimental situations, monopole are counted as 0 when calculating the vertex correlator.
\end{thebibliography}
\end{document}